\documentclass[preprint]{JHEP3}
\usepackage{graphicx}
\newcommand{\alt}{\mathrel{\raisebox{-.6ex}{$\stackrel{\textstyle<}{\sim}$}}}
\newcommand{\agt}{\mathrel{\raisebox{-.6ex}{$\stackrel{\textstyle>}{\sim}$}}}

\title{Triplet Higgs boson at hadron colliders}
\author{Kingman Cheung and Dilip Kumar Ghosh \\
National Center for Theoretical Sciences, National Tsing Hua
University, Hsinchu, Taiwan, R.O.C. \\
Email: cheung@phys.cts.nthu.edu.tw, Dilip.Ghosh@cern.ch
}
\preprint{NSC-NCTS-020828}
\abstract{
The novel feature of a Higgs-triplet representation is a nonzero 
tree-level coupling of $H^+ W^- Z$, which is absent in all 
Higgs-doublet models.  
We study the associated production of a singly-charged Higgs boson 
of the Higgs-triplet representation with a $W$ or $Z$ boson
at hadron colliders, followed by the $H^+ \to W^+ Z$ decay.
We find that the $2\ell+4j$ final state gives an interesting
level of signal with a negligible background, plus it allows a full 
mass reconstruction of the charged-Higgs boson.
The cover range of the charged-Higgs mass is between 110 and 200 GeV.
}

\begin{document}

\section{Introduction}

The Higgs boson is still at large under all present serious efforts in 
searching for it.   The present lower limit on the standard model (SM)
 Higgs boson is 114.1 GeV at 95\% C.L. from LEP Collaborations \cite{lep-sm}.  
The limit on the light neutral-scalar Higgs boson of the minimal 
supersymmetric standard model (MSSM) is 91.0 GeV \cite{lep-mssm}.  
It starts to push into the region that is not so favored by the electroweak
precision measurements \cite{lep-ew}.  
Experimenters should also search for the Higgs-boson
signals other than the usual SM or MSSM Higgs bosons.  
Models with more than one Higgs doublet often predict the existence of
charged-Higgs bosons, which may be singly-, doubly, or even 
triply-charged.
One particularly interesting channel to look for a charged-Higgs boson is
via its coupling to $WZ$.  The reason behind is that if there exists a 
tree-level coupling 
of $H^+ W^- Z$, the Higgs boson must belong to a Higgs structure more
complicated than just doublets (e.g., in MSSM with two Higgs doublets there is
no tree-level $H^+ W^- Z$ coupling.)  A triplet representation is a typical 
example of this kind. 

In extensions of the SM with Higgs doublets and singlets, the coupling
$H^+ W^- Z$ vanishes at tree level and can
only be generated at one-loop level~\cite{GKW,PHI}.
The vanishing of the $H^+ W^- Z$ coupling in the
Lagrangian is due to the hypercharge
($Y$) and weak-isospin assignments of the Higgs representations
introduced in the model. 
In addition, in multi-Higgs-doublet models, the resulting strength of 
the loop-induced $H^+W^-Z$ coupling turns out to be rather small,
of the order of $10^{-2}$
relative to the SM vertex $HW^+W^-$.  Therefore, a large 
$H^+ W^- Z$ coupling is an indicator of  triplet or higher
Higgs representations beyond doublet models.
Therefore, a search for the charged-Higgs boson in the $H^+W^-Z$ channel
would be a test for new physics beyond Higgs-doublet models.

The potential of LEPII and the future TeV $e^+ e^-$ colliders 
in differentiating the 
charged-Higgs bosons of a triplet from a doublet representation had been
studied in Refs. \cite{roger,godbole,ghosh}.
Here we attempt to search for such a singly-charged Higgs boson of the triplet
representation at hadron colliders, with emphasis on the Run II at 
the Tevatron. 
Note that the complex-triplet representations also predict a
doubly-charged Higgs boson ({\em i.e.}~$H^{++}$), which couples to a
pair of same-charged leptons.  Studies of this signature had been done in
Ref.~\cite{VD} for $H^{++}$ at hadron colliders, in Ref.~\cite{e-e-} for
$H^{--}$ production at $e^-e^-$ linear colliders,
and in Ref. \cite{godbole2} for doubly-charged Higgs pair production
at photon colliders.

The organization is as follows.  In the next section, we briefly describe
a viable model of Higgs-triplet representation.  In Sec. III, we highlight
some current bounds on this model.  In Sec. IV, we calculate the associated 
production of the charged-Higgs boson with a $W$ or $Z$ boson, and 
discuss the possible signatures over the backgrounds.  We conclude in Sec. V.

\section{The Model}

Here we consider the triplet-Higgs model 
by Galison~\cite{PG}, and by Georgi and Machacek~\cite{GM}.
They introduced more than one Higgs-triplet field into
the model and imposed an $SU(2)$ custodial symmetry on the vacuum
expectation values and hypercharges of the Higgs multiplets
to ensure $\rho=1$ at tree level.  
Stability conditions of the $SU(2)$ custodial symmetry in the Higgs 
potential under higher-order quantum corrections were 
further analyzed by Chanowitz and Golden~\cite{CG}.
The model consists of a SM $Y=1$ complex doublet $\Phi$, plus one real 
$Y=0$ triplet and one complex $Y=2$ triplet. 
We follow closely the convention of Ref. \cite{jack}
\begin{equation}
\phi = \left( \begin{array}{cc}
\phi^{0^*} & \phi^+ \\
\phi^-     &  \phi^0  \end{array} \right )\;, \qquad
\Delta\ =\ \left( \begin{array}{ccc}
\chi^0 & \xi^+ & \chi^{++} \\
\chi^- & \xi^0 & \chi^+ \\
\chi^{--} & \xi^- & \chi^{0\ast} \end{array} \right).
\end{equation}
The tree-level gauge-boson masses are fixed by the kinetic energy terms of 
the Higgs bosons, which are
\[
{\cal L} = \frac{1}{2} \, {\rm Tr} \left [ 
(D_\mu \phi)^\dagger (D^\mu \phi) \right ] 
+ \frac{1}{2} \, {\rm Tr} \left [ 
(D_\mu \chi)^\dagger (D^\mu \chi) \right ] \;,
\]
where $D_\mu \phi$ and $D_\mu \chi$ are the covariant derivatives taking 
into account the SU(2) in $2\times 2$ and $3\times 3$ representations, 
respectively. 
In order to preserve $\rho=1$, a custodial SU(2)$_R$ symmetry is imposed 
such that the Lagrangian is invariant under the global SU(2)$_L\times$ 
SU(2)$_R$ symmetry.
In particular, the tree-level invariance of the gauge-boson mass terms under
this custodial SU(2)$_R$ symmetry is arranged by giving $\chi^0$ and 
$\xi^0$ the same vacuum expectation value (VEV).  
The VEVs for the fields are defined as
\[
\langle \phi^0 \rangle= \frac{v_D }{\sqrt 2}\,, \qquad
\langle \chi^0 \rangle= \langle \xi^0 \rangle = v_T \;.
\]
It is also convenient to define 
\begin{equation}
v^2 = v_D^2 + 8 v_T^2 \;, \qquad 
\sin \theta_H = \frac{ \sqrt{8} v_T} { \sqrt{ v_D^2 + 8 v^2_T } }\;, \qquad
\cos \theta_H = \frac{ v_D} { \sqrt{ v_D^2 + 8 v^2_T } } \;,
\end{equation}
where $\theta_H$ is the doublet-triplet mixing angle.   

By absorbing the Goldstone bosons the $W$ and $Z$ bosons acquire 
masses given by $m_W = g v /2$ and $m_Z = m_W/\cos^2 \theta_{\rm w}$.  
The original number of degrees of freedom in the Higgs sector is 13
(1 complex doublet, 1 real triplet, and 1 complex triplet).  Therefore,
after 3 degrees of freedom become the longitudinal components of the gauge 
bosons, there are 10 physical states, which form one 
five-plet, one three-plet, and two singlets:
\begin{equation}
H_5 = ( H^{++}_5,\, H^{+}_5,\, H^{0}_5,\, H^{-}_5,\, H^{--}_5 )^T\;, \qquad
H_3 = ( H^{+}_3,\, H^{0}_3,\, H^{-}_3 )^T\;, 
H_1^0,\, H_1^{0'} \;,
\end{equation}
which are given in terms of the original fields as
\begin{eqnarray}
H_5^{++} &=& \chi^{++}\,, \;\; 
H_5^{+} = \frac{1}{\sqrt 2} ( \chi^+ - \xi^+ ) \,, \;\;
H_5^0     = \frac{1}{\sqrt 6} ( 2 \xi^0 - \sqrt{2} \chi^0 )\,, \nonumber \\
H_3^+ &=& \frac{\cos \theta_H }{\sqrt 2} ( \chi^+ + \xi^+) -
   \sin\theta_H \phi^+ \;, \;\;
H_3^0 = i ( \cos\theta_H \chi^{0i} + \sin \theta_H \phi^{0i} ) \;, \nonumber \\
H^0_1 &=& \phi^{0r} \,, \;\; 
H_1^{0'} = \frac{1}{\sqrt 3} ( \sqrt{2} \chi^{0r} + \xi^0 ) \;.
\end{eqnarray}
These Higgs fields can also mix.  However, if the custodial SU(2) symmetry 
is preserved in the Higgs potential,  the five-plet and three-plet will
not mix with each other or with the singlets.  The only possible
mixing is between
the two singlets.  For simplicity we assume no further mixing of the above 
states and so they are the physical Higgs states.  

Phenomenology is mainly determined by the Higgs couplings to fermions
and gauge bosons.  Recalling that the standard Yukawa coupling is via
the doublet-Higgs fields to the fermion-antifermion pair, the 
coupling of a Higgs state to a fermion-antifermion pair is determined by its
doublet component.  Thus, the whole $H_5$ and $H_1^{0'}$ have no
fermion-antifermion couplings.  The only fermionic coupling of $H_5$ is
the $H_5^{++} \ell^- \ell^-$ coupling, which is not present in the SM.
Other than that, only $H_3$ and $H_1^0$ have fermionic couplings. 
On the other hand, $H_3$ has no tree-level coupling to gauge bosons while
all the others have.
A novel feature is the existence of a nonzero tree-level coupling 
of $H_5^+$ to $W^- Z$, which is absent in all Higgs-doublet models.  
The observation of such a coupling necessarily signals a Higgs structure
more complicated than doublets.   This is the main motivation of the present
work.  

The corresponding vertex $H_5^+ W^-Z$ is given by~\cite{jack}
\begin{equation}
{\cal L}_{int} \ =\ -g \frac{\sin\theta_H}{\cos\theta_{\rm w}} \,
   M_W H^+_5 W^{-\mu} Z_\mu\ +\ h.c.,
\label{int}
\end{equation} %4
where $g$ is the usual $SU(2)_L$ electroweak coupling constant,
$\cos^2\theta_{\rm w}=1-\sin^2\theta_{\rm w}=M^2_W/M^2_Z$ at tree-level,
and $\theta_{\rm w}$ is the Weinberg angle.   
Due to the electromagnetic gauge invariance,
the coupling $H_5^+ W^- \gamma$ is absent at tree level. As emphasized
earlier, we are interested in a large $H_5^+ W^- Z$ coupling that will
unavoidably signify the triplet nature of the charged-Higgs boson $H_5^+$.
However, this can only be possible experimentally 
if $\sin\theta_H \sim 1$ or equivalently
$v_T \sim v_D$, which is considered to be a natural scenario.
The partial width of $H^+_5 \to W^+ Z$ is given by
\begin{equation}
\label{GH}
\Gamma (H^+_5 \to W^+ Z) = \frac{\alpha_w \sin^2\theta_H} {16} 
\; M_{H_5} \lambda^{1/2} \left( \frac{M^2_{H_5}}{M^2_W},\frac{1}{
\cos^2\theta_{\rm w}},1 \right ) \,
\Big[ 1+ x_W^2 + x_Z^2 -2 x_W -2 x_Z +10 x_W x_Z \Big] \;
\end{equation}
with $\alpha_w=g^2/4\pi$, $\lambda(x,y,z)=(x-y-z)^2-4yz$, 
$x_W=M_W^2/M_{H_5}^2$, and $x_Z=M_Z^2/M_{H_5}^2$.

There are other triple vertices that will affect the decay of the $H_5^+$.
They are $H_5^+ H_3^0 W^-$ and $H_5^+ H_3^- Z$, the vertex factors of which
are given by  $-ig \cos\theta_H (p-p')^\mu/2$ and 
$ig \cos\theta_H (p-p')^\mu/(2 \cos\theta_{\rm w})$, as well as
$H_5^+ H_3^- H_3^0$, which depends
crucially on the details of the Higgs potential. 
Therefore, in general $H_5^+$ can decay into $W^+ Z, H_3^+ Z, H_3^0 W^+$, and
$H_3^+ H_3^0$, some of which may be off-shell because of kinematics.
(Note that all five members of the five-plet are of the same mass because of
the custodial SU(2) symmetry, and so are the three members of the
three-plets.  Hence, we do not have to consider the decay of $H_5^+$ into 
any other members of the five-plet.)
Our main interest is the $W^+ Z$ decay channel of the $H_5^+$, which can be 
achieved by some not-so-fine tuning of the parameters of the model.  The 
simplest approach is to make three-plet heavier than the five-plet.  
The masses for the five-plet and the three-plet are given by
\begin{equation}
\label{mass}
m^2_{H_5} = 3 ( \lambda_5 \sin^2\theta_H + \lambda_4 \cos^2 \theta_H ) \, v^2
\;, \qquad
m^2_{H_3} = \lambda_4 \, v^2 \;,
\end{equation}
where $\lambda_4$ and $\lambda_5$ are the parameters in the Higgs potential.
It is obvious that if we put $\cos^2\theta_H \le 0.3$ and $\lambda_5 \ll 
\lambda_4$, then $M_{H_5} < M_{H_3}$.  Thus, $H_5^+$ decays dominantly into
$W^+ Z$.   This corresponds to $\tan\theta_H \agt 1.5$.  
In the next section, we shall highlight the existing limits on $\tan\theta_H$,
and we shall see that such a $\tan\theta_H$ range is still allowed.

On the other hand, if $\tan\theta_H < 1.5$ then $M_{H_5} > M_{H_3}$ and 
the decay modes $H_5^+ \to H_3^+ Z, H_3^0 W^+$, and $H_3^0 H_3^+$ open
up. In fact, they could be dominant for a very small $\tan\theta_H$
\cite{andrew}.  Since $H_3$ decays mainly into a fermion-antifermion pair,
the strategy for searching for $H_5^+$ changes somewhat \cite{future}.  
An exhaustive list of the Feynman rules containing all the Higgs particles
involved in this model can be found in Ref. \cite{jack}.

\section{Review of bounds on $\tan \theta_H$}

A number of low-energy precision measurements constrain the Higgs-triplet 
model, e.g., $Z\to b\bar b$ vertex, $B^0 - \overline{B^0}$ mixing, 
$K^0 - \overline{K^0}$ mixing, and the ratio of $b\to u$ to $b\to c$ decays:
for a summary see Refs. \cite{jack,logan}.  The strongest bound comes from
the $Z\to b\bar b$ vertex with the charged-Higgs boson in the loop 
\cite{kundu2,ciu,logan}.  
Note that if the Higgs potential satisfies the custodial SU(2) symmetry, 
which is preferred in order to fulfill $\rho=1$, the five-plet and the
three-plet do not mix, and thus only the three-plet couples to the
fermion-antifermion pair.  Hence, all these bounds are put directly on
$M_{H_3}$ and $\tan\theta_H$.  In general, when $M_{H_3}$ gets larger, the
bound on $\tan\theta_H$ will be relaxed.  However, unitarity
requires $M_{H_3}$ not to be larger than about 1 TeV, otherwise, the 
longitudinal-$W$-boson scattering becomes so strong that unitarity 
would be violated. We are going to summarize the existing bounds.

An early analysis in Ref. \cite{jack} used $\epsilon_K$, 
$B^0 - \overline{B^0}$ mixing, and the ratio of $b\to u$ to $b\to c$ decays.
The acceptable range is either $M_{H_3} \agt 1$ TeV with $\tan\theta_H > 5$, or
$\tan\theta_H \alt 1.5$, which took into account reasonable variations
on hadronic uncertainties. Reference \cite{kundu1} refined the analysis on
meson mixings and obtained the bound $\tan\theta_H < 6-7 \;(3-3.5)\; (1-1.2)\;$
for $M_{H_3}=100-500$ GeV.  These three ranges are for different hadronic
inputs and  CKM phases.  Despite all these bounds the strongest comes from
$Z\to b\bar b$ vertex \cite{kundu2,ciu,logan}. Kundu and Mukhopadhyaya
\cite{kundu2} obtained a bound $\sin\theta_H > 0.8$ for $M_{H_3} \alt 1$ TeV.
An NLO analysis in MSSM \cite{ciu} obtained $\tan\beta > 1.8, 1.4, 1.0$ for 
$M_{H^+}=85, 200, 425$ GeV, which are equivalent to $\tan\theta_H < 0.555,
0.71, 1$.  The most updated analysis comes from Haber and Logan
\cite{logan}.  The limits at 95\% C.L. are $\tan\theta_H \alt 0.5, 1, 1.7$
for $M_{H_3}=0.1, 0.5, 1$ TeV.

Our main interest is the $W^+Z$ decay mode of $H_5^+$.  In order to 
validate this scenario, we choose $M_{H_3}$ to be very heavy ($\alt 1$ TeV
but it is not important as long as $M_{H_5}< M_{H_3}$) and $\tan\theta_H
\alt 2$.  Once $\tan\theta_H \agt 1.5$,  $M_{H_5}$ can be chosen to be lighter
than $M_{H_3}$ by tuning the parameters $\lambda_4$ and $\lambda_5$ in
Eq. (\ref{mass}).  Although a somewhat-fine tuning ($\lambda_5<0,
\lambda_4>0$) is needed to make $M_{H_5}$ to be of order of $100-200$ GeV,
there are no obvious constraints on $\lambda_4$ and $\lambda_5$ from current
experiments.  
In the following, we choose $\tan\theta_H=1-2$ as typical inputs and we 
are interested in $M_{H_5}=100-200$ GeV for an observable cross section
at the Tevatron.

\section{Production at the Tevatron}

The processes for the associated production of $H^+_5$ ($H_5^-$) 
with a $W$ or $Z$ boson in hadronic collisions are
\begin{eqnarray}
q \bar q &\to & W^- H_5^+ \;\; (W^+ H_5^-)\\
q \bar q' &\to & Z H_5^+ \;\; (Z H_5^-)\\
q  q' &\to& q q^{''} W^* Z^* \to q q^{''} H_5^+ \;\; (q q^{''} H_5^-) \;.
\end{eqnarray}
We have included the charged-conjugated processes in our analysis.  The 
first two processes are the Higgs-bremsstrahlung off the $W$ or $Z$, while 
the last one is the $WZ$ fusion.   The last process is sub-dominant at the
energy of the Tevatron, but will dominate at the LHC instead \cite{future}.
For the present work, we shall ignore the last process in our study.  

The leading-order (LO) subprocess cross sections for $VH_5^\pm$ are given by
%\begin{eqnarray} \label{cs}
%\hat{\sigma}_{\!\mbox{\tiny LO}}(q\bar{q^\prime} \to V H_5^\pm )
%&=& \frac{G_{\scriptscriptstyle F}^2 \, M_{\scriptscriptstyle V}^4}{288 \,\pi\,
% \hat s}(v_q^2 +a_q^2) \;
%\frac{\lambda(1,\,
% m_{\scriptscriptstyle V}^2/\hat s, \, m_{\scriptscriptstyle {H_5}}^2/
%  \hat s)+12 \,m_{\scriptscriptstyle V}^2/\hat s}
%{(1-m_{\scriptscriptstyle V}^2/\hat s)^2} \nonumber \\
%&&\times 
%\sqrt{\lambda(1,\,m_{\scriptscriptstyle V}^2/\hat s,\,
%   m_{\scriptscriptstyle {H_5}}^2/\hat s)}\; \sin^2\theta_H
%\end{eqnarray}
%where $v_q=-a_q=\sqrt{2}$ for $V=Z$ ($V'=W$) and 
%$v_q= 2 \, T_{3q} - 4 \,Q_q \, \sin^2\theta_{\rm w}$, \ $ a_q=2\ T_{3q}$ for 
%$V=W $ ($V'=Z,\; q^\prime =q$), and 
%$\lambda(x,y,z)=(x-y-z)^2-4yz$.
%Note that when $V=W(Z)$ in the final state, the $V'= Z (W)$ in the 
%propagator.
\begin{eqnarray} \label{cs}
\hat{\sigma}_{\!\mbox{\tiny LO}}(q\bar{q^\prime} \to W^* \to Z H_5^+ )
&=& \frac{G_{\scriptscriptstyle F}^2 \, m_{\scriptscriptstyle W}^4}{72 \,\pi\,
 \hat s} \;
\frac{\lambda(1,\,
 m_{\scriptscriptstyle Z}^2/\hat s, \, m_{\scriptscriptstyle {H_5}}^2/
  \hat s)+12 \,m_{\scriptscriptstyle Z}^2/\hat s}
{(1-m_{\scriptscriptstyle W}^2/\hat s)^2} \nonumber \\
&&\times 
\sqrt{\lambda(1,\,m_{\scriptscriptstyle Z}^2/\hat s,\,
   m_{\scriptscriptstyle {H_5}}^2/\hat s)}\; \sin^2\theta_H \\
\hat{\sigma}_{\!\mbox{\tiny LO}}(q\bar{q} \to Z^* \to W^- H_5^+ )
&=& \frac{G_{\scriptscriptstyle F}^2 \, m_{\scriptscriptstyle W}^4}{36 \,\pi\,
 \cos^2 \theta_{\rm w} \hat s} \; ( g_L^2 + g_R^2 )
\frac{\lambda(1,\,
 m_{\scriptscriptstyle W}^2/\hat s, \, m_{\scriptscriptstyle {H_5}}^2/
  \hat s)+12 \,m_{\scriptscriptstyle W}^2/\hat s}
{(1-m_{\scriptscriptstyle Z}^2/\hat s)^2} \nonumber \\
&&\times 
\sqrt{\lambda(1,\,m_{\scriptscriptstyle W}^2/\hat s,\,
   m_{\scriptscriptstyle {H_5}}^2/\hat s)}\; \sin^2\theta_H \nonumber 
\end{eqnarray}
where $g_L(q) = T_{3q} -Q_q \sin^2 \theta_{\rm w}$ and 
      $g_R(q) =        -Q_q \sin^2 \theta_{\rm w}$, and 
      $\lambda(x,y,z)=(x-y-z)^2-4yz$.

\begin{figure}[th!]
\includegraphics[width=5in]{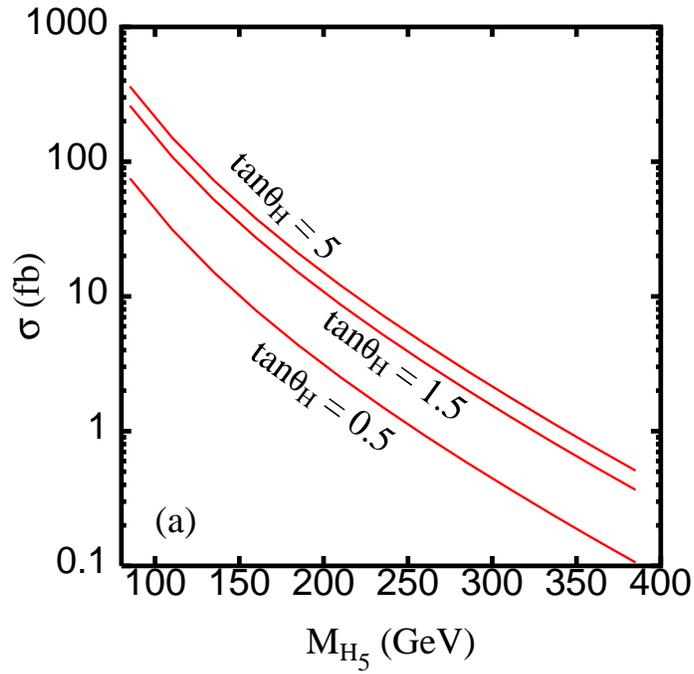}

\includegraphics[width=5in]{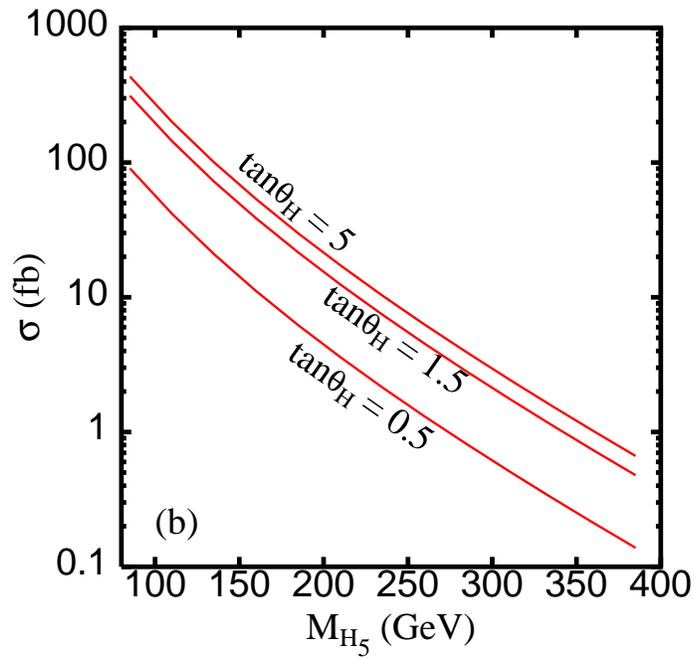}
\caption{\small
\label{total}
Total production cross sections for (a) $p\bar p \to 
W^+ H_5^- + W^- H_5^+$ and (b) $Z H_5^+ + Z H_5^-$ for $\tan\theta_H=0.5,1.5,5$
at $\sqrt{s}=2$ TeV.}
\end{figure}

The subprocess cross sections are then convoluted with the parton distribution
functions to obtain the total production cross sections.  Throughout our
analysis we use the CTEQ5L distribution set \cite{cteq}.
In Fig. \ref{total}, we show the total production cross sections for 
$W^+ H_5^- + W^- H_5^+$ (a) and $Z H_5^+ + Z H_5^-$ (b) as a function of
$M_{H_5}$ for three choices of $\tan \theta_H=0.5,\,1.5$, and $5$. 
The cross section increases substantially from $\tan\theta_H=0.5$ to $1.5$,
but only slightly from $\tan\theta_H=1.5$ to $5$. 
This can be easily understood by the explicit dependence on $\sin \theta_H$,
as shown in Eq. (\ref{cs}).

The next concern in our analysis is the decay channels and various backgrounds.
Since the number of combinations in the decays are quite complicated, we 
will demonstrate with the best decay channel and the corresponding 
backgrounds.

Since we are mainly interested in the $W^+Z$ decay mode of $H_5^+$ and we want
to have a fully-reconstructed Higgs mass, we would choose the following decay
mode of the $H_5^+$
\begin{equation}
\label{444}
H_5^+ \to W^+\; Z \to (q \bar q') \; (\ell^+ \ell^-) \;.
\end{equation}
We have assumed that $H_5^+ \to W^+ Z$ with a 100\% branching ratio,
which is made possible by adjusting the parameters
$\lambda_4$ and $\lambda_5$ of Eq. (\ref{mass}).
The combined branching ratio for the channel in Eq. (\ref{444}) 
is about $0.7 \times 0.068=0.048$, which takes into account
both the electron and muon modes of the $Z$ decay.  The branching ratio would 
increase if we chose the hadronic mode of the $Z$, but it would make the
jet combinatorics too complicated for a clean reconstruction.   The decay
mode of the associated $W$ or $Z$ boson can be either leptonic or hadronic.
Therefore, we have the following modes in the final state
\begin{equation}
W H_5^\pm \to W \; (WZ^*/W^* Z) \to \left \{\begin{array}{c}
             (\ell\nu)\; (q\bar q')\; (\ell^+\ell^-) \\
             (q\bar q')\; (q\bar q')\; (\ell^+\ell^-)
                                       \end{array} \right. \;,
\end{equation}
and
\begin{equation}
Z H_5^\pm \to Z \; (WZ^*/W^* Z) \to  \left \{\begin{array}{c}
          (\ell^+\ell^-)\; (q\bar q')\; (\ell^+\ell^-)\\
         (q\bar q)\; (q\bar q')\; (\ell^+\ell^-)
                                       \end{array} \right . \;,
\end{equation}
where ``*'' denotes an offshell vector boson.  These channels result in 
$3\ell+2j +\not{\!E}_T$, $2\ell+4j$, or $4\ell+2j$ in the final state.
The signal is a $\alpha_w^2$ process.

The irreducible backgrounds come from \cite{han}
\begin{equation}
p\bar p \to W^+ W^- Z,\; Z W^\pm Z,\; ZZZ \;,
\end{equation}
which are $\alpha_w^3$ processes.  Thus, before imposing any cuts, these
backgrounds are already subdominant relative to the signal.  
The cross sections for $W^+W^-Z$, $Z W^\pm Z$, and $ZZZ$ at the 2 TeV 
Tevatron are $6.2$, $1.6$ and $0.6$ fb, respectively.
We therefore 
do not impose specific cuts to suppress these backgrounds, except for
the selection cuts for leptons and jets.

Other reducible backgrounds include $W+$jets, $Z+$jets, $WW+$jets, $ZZ+$jets,
and $WZ+$jets \cite{multi-W}.  
The $V+$jets ($V=W,Z$) are $\alpha_w \alpha_s^n$ processes
whose cross sections can be, in principle, larger than the signal 
cross sections.   However, they can be reduced substantially by imposing
a transverse momentum ($p_T$) cut on the jets and by requiring a pair
of the jets reconstructed at the $W$ or $Z$ mass.  Note that the jets of
the signal that come off from the $H_5^+$ decay have a
relatively much larger $p_T$.
The $VV+$jets ($V=W,Z$) are $\alpha_w^2 \alpha_s^n$ processes, which are
already suppressed relative to the signal.  

Next we describe our analysis in details.  We apply a typcial resolution
\cite{tev2000}
\begin{equation}
\frac{\Delta E}{E} = \frac{0.3}{\sqrt{E}} \oplus 0.01
\end{equation}
for leptons, where $E$ is in GeV, and 
\begin{equation}
\label{jet-res}
\frac{\Delta E}{E} = \frac{0.8}{\sqrt{E}} \oplus 0.05
\end{equation}
for jets.
We impose the following selection cuts on leptons and jets \cite{tev2000}
\begin{eqnarray}
p_T(\ell) &>& 10\; {\rm GeV} \;, \nonumber\\
p_T(j) &>& 15\; {\rm GeV}  \;,\nonumber\\
\left | \eta_\ell \right | &<& 2.5 \;,\nonumber \\
\left | \eta_j \right | &<& 2.5 \;,\nonumber \\
\Delta R_{jj} &>& 0.7 \;,\nonumber \\
\Delta R_{\ell j} &>& 0.4 \;,\nonumber  \\
\not{\!E}_T &>& 20 \;{\rm GeV} \;\; 
\mbox{when considering $W\to \ell\nu$} \nonumber \;.
\end{eqnarray}

{}From the above discussion we see that the final state of the signal 
consists of $3\ell+2j +\not{\!E}_T$, $2\ell+4j$, or $4\ell+2j$.  
Let us first concentrate on the $2\ell+4j$ mode because of its largest
branching ratio.  We shall comment on the other two modes later.
In the $2\ell+4j$ mode, there are a few 
combinations to determine the $2\ell + 2j$ that decay from the $H_5^+$.  
We employ the following procedures to select the right combination.

\begin{figure}[th!]
\includegraphics[width=5in]{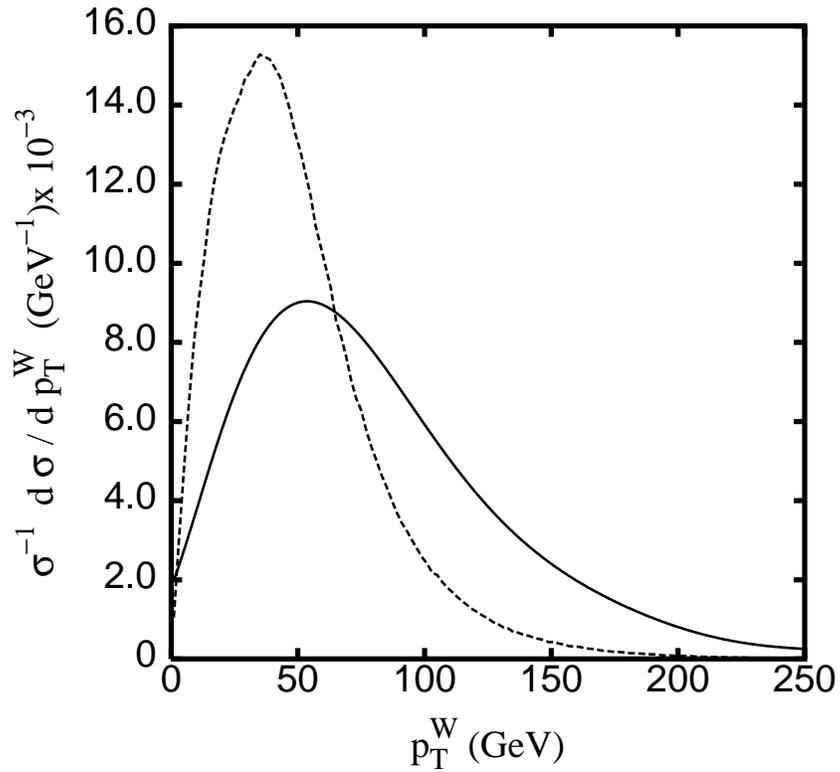}
\caption{\small
\label{ptw}
Normalized transverse-momentum spectra for the associated $W$ (solid) 
and for the $W$ (dashed) decaying from $H_5^+$ in the production of 
$p\bar p \to W^\mp H_5^\pm \to W^\mp W^\pm Z$ at the 2 TeV Tevatron.
}
\end{figure}

\begin{figure}[th!]
\includegraphics[width=5in]{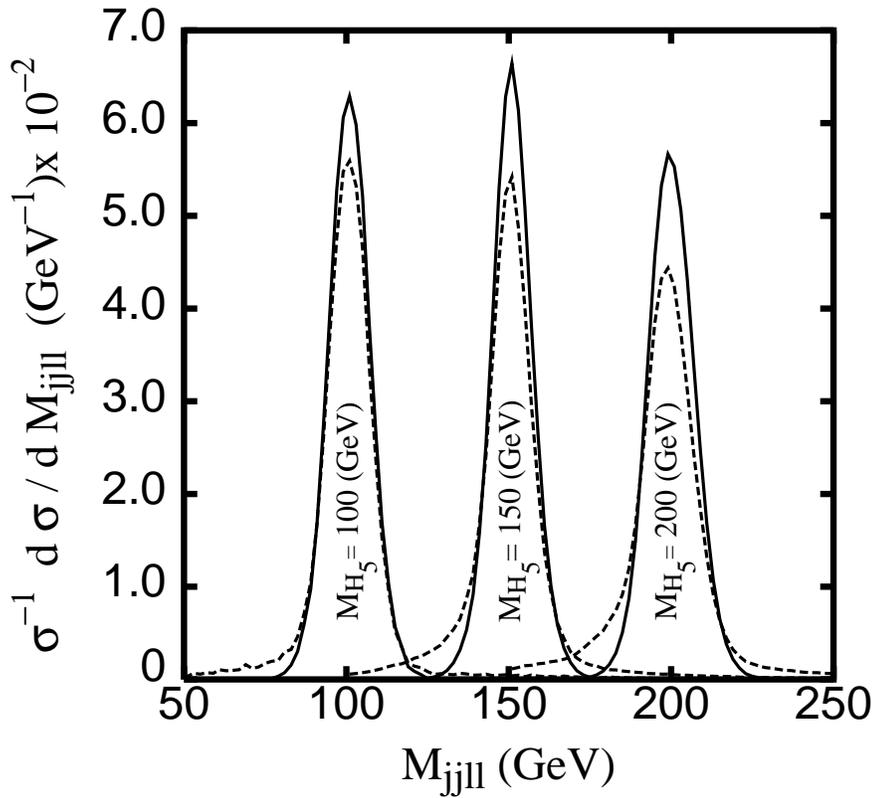}
\caption{\small
\label{mh-recon}
Normalized invariant-mass distributions for various $H_5^+$ masses in the 
production of $p\bar p \to W^\mp H_5^\pm + Z H_5^\pm$ at 
the 2 TeV Tevatron.  Both the theoretical peaks (solid) and the 
reconstructed peaks (dashed) are shown for comparison.
}
\end{figure}

First, we reconstruct the associated $W$ or $Z$ boson by demanding 
that 
\begin{eqnarray}
| M_{jj} - M_Z | &<& 10 \;{\rm GeV}\,, \;\;\; \nonumber \\
| M_{jj} - M_W | &<& 10 \;{\rm GeV} \,,\;\;\; {\rm or}\;\;\; \nonumber \\
| M_{\ell\ell} - M_Z | &<& 10 \; {\rm GeV}\,, \label{jj}
\end{eqnarray}
where the $2j$ can come from a $W$ or a $Z$ boson while $2\ell$ can only
come from the $Z$ boson.  
It could happen that more than one jet or lepton pair satisfy
Eq. (\ref{jj}). In this case, we choose the pair 
that has a higher transverse momentum $p_T$, because we expect that 
the associated $W$ or $Z$ boson has a higher
$p_T$ than the boson decaying from the $H^+_5$. For illustration 
we show in Fig. \ref{ptw}
the normalized transverse momentum distributions for the associated 
$W$ and for the $W$ 
decaying from $H_5^\pm$ in the $W^\mp H_5^\pm$ production.  From the figure 
it is easy to see that when we 
select the reconstructed vector boson with a higher transverse momentum, we
are more likely to pick the correct associated vector boson.  Once
we select the correct associated vector boson, we can then reconstruct the
invariant mass of the other particles in the final state to form
the $H_5^+$.  In Fig. \ref{mh-recon}, we show both the theoretical $H_5^+$
mass peaks and the peaks formed by the above procedures. It is clear that
our procedures can select the right combination most of the time.

\begin{figure}[th!]
\includegraphics[width=5in]{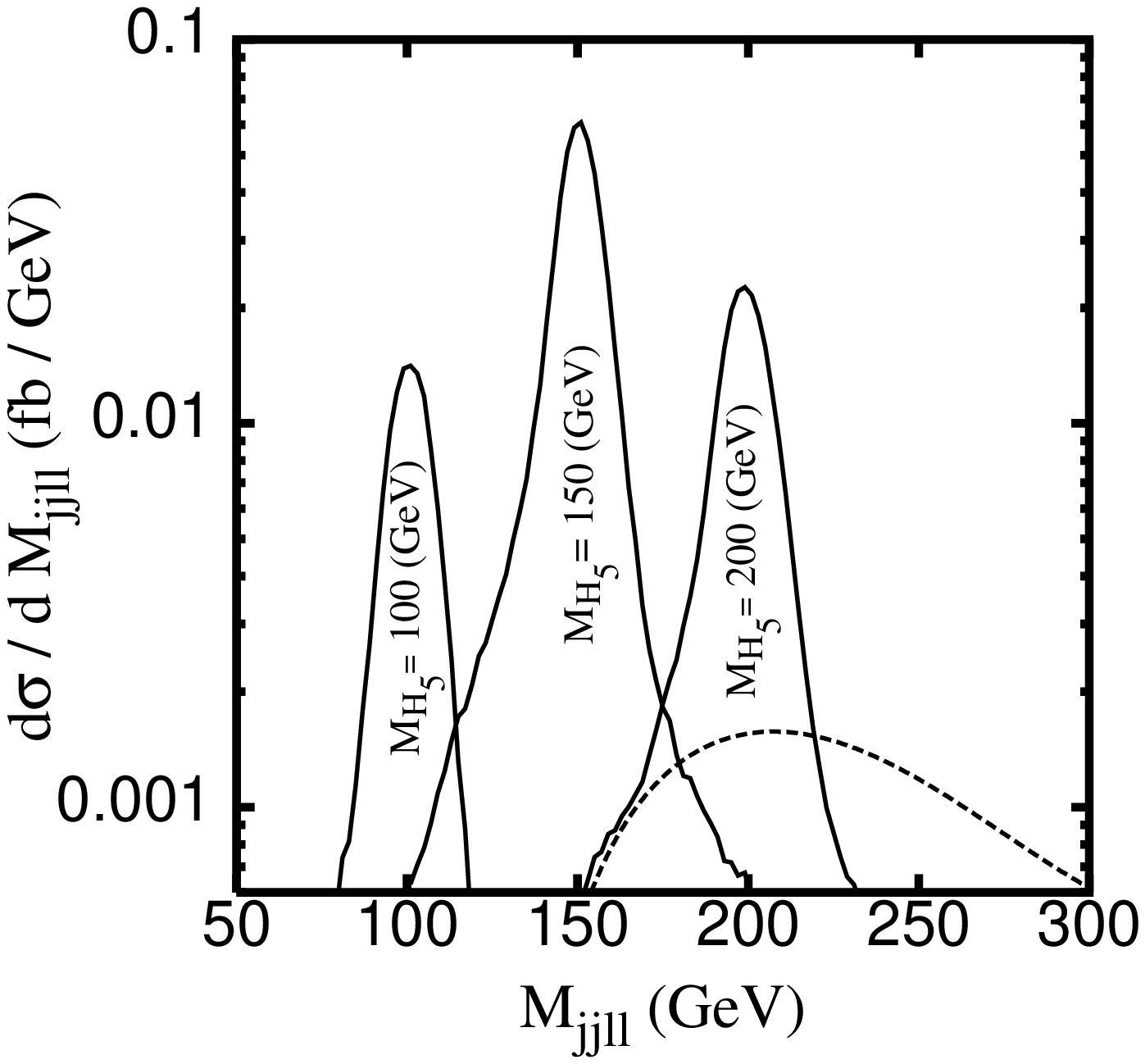}
\caption{\small
\label{m-jjll}
Differential cross section $d\sigma/d M_{jj\ell\ell}$ versus 
$M_{jj\ell\ell}$ in $p\bar p$ collisions at the 2 TeV Tevatron.
We show the signal of $W^\pm H_5^\mp + Z H_5^\pm \to (jj) (jj) (\ell\ell)$ 
for $M_{H_5}=100, 150, 200$ GeV and $\tan\theta_H=1.5$. The continuum
background (dashed) from $p\bar p \to WWZ, ZWZ, ZZZ$ is also shown.}
\end{figure}

We apply exactly the same procedures to the backgrounds. In the backgrounds,
all three vector bosons are on equal footing.  Our reconstruction procedures
will not select any particular one.  The reconstructed spectrum
of $jj\ell\ell$ would not show any peak structures but a continuum.  
In Fig. \ref{m-jjll}, we show both the background spectrum and the signal
peaks for various $H_5^+$ masses.  The background spectrum
includes contributions from $WWZ, ZWZ$, and $ZZZ$.  Similarly, 
we have added contributions from $W^\pm H_5^\mp$ and $Z H_5^\pm$ in the signal.
It is obvious that the background is almost negligible under the Higgs peaks.
Therefore, the criteria for a discovery of $H_5^+$ depends crucially
on the number of signal events.  We require a minimum of 5 events for the
evidence of existence.  In Table \ref{tab},
we show the signal cross sections in fb for the signals of
$W^\pm H_5^\mp + Z H_5^\pm$ in the $2\ell+4j$ mode.  Given an integrated
luminosity of 20 fb$^{-1}$ accumulated in the Run 2b of the Tevatron, the
sensitive range of $H_5^+$ is between 110 and 200 GeV.  The cross section for
$M_{H_5^+} \approx 100$ GeV is small because of the $p_T$ cuts on the leptons
and jets that decay from the $H_5^+$.  The heavier the Higgs boson the larger
is the $p_T$ of its decay products. 

\section{conclusions}

The spectacular signal of the existence of $H_5^\pm$ is its decay into a $WZ$ 
pair.  We chose the $H_5^\pm$ decay mode $(q\bar q' \ell^+ \ell^-)$ for a clean
reconstruction of the $WZ$ pair.  
Had we chosen the $4j$ mode, it would
have been very difficult to be identified as a $WZ$ pair.
The $2\ell+4j$ final state gives an interesting level of signal event rates
with a negligible background.  A minimum requirement of 5 signal events
allows the possible evidence of existence of $H_5^\pm$ between 110 and 200 GeV.

In our analysis, we have not taken into account the QCD corrections to the
signal and backgrounds.  The QCD correction to the standard model
$VH$ production was 
known to be about 40\% at the Tevatron \cite{spira}.  We expect about the
same enhancement to the $V H_5^\pm$ and $VVV$ production as the 
QCD correction is 
independent of the final states.  Therefore, the observability of the $H_5^\pm$
signal improves, may be up to about 210 GeV.

The decay mode of $H_5^+$, $H_5^+ \to W^+ Z \to (q \bar q') (\ell\ell)$,
might be mimicked by the SM Higgs signal, $H_{\rm SM} \to ZZ \to
(q\bar q) (\ell\ell)$.  Nevertheless, the mass reconstruction of the $W$ boson
can help us to distinguish the triplet-Higgs signal from the SM one.
The jet resolution given in Eq. (\ref{jet-res}) is good enough to provide
a reasonable $W$-boson identification.   Suppose the $W$ boson decays into
2 jets, each of which has an energy about 50 GeV. According to 
Eq. (\ref{jet-res}), the $\Delta E$ of each jet is then about 6 GeV. 
Thus, the $M_{jj}$ mass resolution is about 8.5 GeV, which 
is better than the mass difference between 
the $W$ and the $Z$ bosons.  

The other two decay modes $3\ell+2j+\not{\!E}_T$ and $4\ell+2j$ 
would result in an
even smaller event rate.  That was the reason why we did not pursue it 
further.  Finally, the $WH_5^\pm \to W\,(W^\pm Z) \to (\ell\nu)(jjjj)$ mode 
suffers from immense background from $t\bar t$ production.

In our analysis, 
the background estimation is based on the on-shell production approximation.
If we had taken the vector bosons off-shell, there would have been
 a small tail at
the small invariant mass region in the background curve in Fig. \ref{m-jjll}.
However, this tail is suppressed by $\alpha_w$ relative to the on-shell 
production.  Thus, it is negligible compared to the Higgs signal peaks.

There are other continuum 
backgrounds that we have not taken into account, e.g.,
$VV+$jets and $V+$jets.  We believe that they are suppressed, as we have 
mentioned earlier,  by our cuts to a level
even smaller than the background that we have considered in this work.
\footnote
{There are some estimates of $VV+$jets background in MadEvents \cite{madevent},
but we used different cuts.
}

\section*{Acknowledgment}
This research was supported in part by the National Center for Theoretical
Science under a grant from the National Science Council of Taiwan R.O.C.

%%%%%%%%%%%%%%%%%%%%%%%%%

%%%%%%%%%%%%%%%%%%%%%%%
\begin{table}
\caption{\small \label{tab}
Signal cross sections in fb for $p\bar p \to W^\pm H_5^\mp,
Z H_5^\pm $ in the $2\ell+4j$ mode.  
Cuts and branching ratios have already been 
included in the cross sections.  Event rates are also shown for an integrated
luminosity of 20 fb$^{-1}$.
}
\medskip
%\begin{ruledtabular}
\begin{tabular}{ccccc}
\hline
\hline
 $M_{H_5^+}$ (GeV) & $\sigma(W^\pm H_5^\mp)$  (fb)& 
  $\sigma(Z H_5^\pm$) (fb) & $\sigma(W^\pm H_5^\mp+Z H_5^\pm)$ (fb) & 
  Signal events  \\
\hline
  100  & 0.05 &   0.08 &   0.13 &  2.6   \\
  110  & 0.15 &   0.22 &   0.37 &  7.4 \\
  120  & 0.22 &   0.30 &   0.51 &  10  \\ 
  140  & 0.25 &   0.35 &   0.60 &  12  \\
  160  & 0.20 &   0.29 &   0.50 &  10  \\
  180  & 0.14 &   0.25 &   0.38 &  7.6  \\
  200  & 0.09 &   0.16 &   0.25 &  5.0 \\
  210  & 0.07 &   0.13 &   0.20 &  4.0 \\
\hline
\end{tabular}
%\end{ruledtabular}
\end{table}

\newpage

\begin{thebibliography}{99}

\bibitem{lep-sm}
LEP Higgs Working Group, hep-ex/0107029. 

\bibitem{lep-mssm}
LEP Higgs Working Group, hep-ex/0107030.

\bibitem{lep-ew}
LEP Electroweak Working Group, LEPEWWG/2002-01.

\bibitem{GKW} J. Gunion, G. Kane, and J. Wudka, Nucl. Phys. {\bf B299}, 231
(1988).

\bibitem{PHI} M. Peyran\`ere, H. Haber, and P.Irulegui,
Phys. Rev. {\bf D44}, 191 (1991).

\bibitem{roger}
K. Cheung, R. Phillips, and A. Pilaftsis, Phys.Rev. {\bf D51}, 4731 (1995).

\bibitem{godbole}
R.M. Godbole, B. Mukhopadhyaya, and M. Nowakowski, Phys.Lett. {\bf B352}, 388
(1995).

\bibitem{ghosh}
D.K. Ghosh, R.M. Godbole, and B. Mukhopadhyaya, Phys. Rev. {\bf D55}, 
3150 (1997).

\bibitem{VD} R. Vega and D. Dicus, Nucl. Phys. {\bf B329}, 533 (1990).

\bibitem{e-e-} V. Barger, J. Beacom, K. Cheung, and T. Han, 
Phys. Rev. {\bf D50},6704 (1995);
R. Alanakian, Phys. Lett. {\bf B436}, 139 (1998).

\bibitem{godbole2}
S. Chakrabarti, D. Choudhury, R.M. Godbole, and B. Mukhopadhyaya,
Phys. Lett. {\bf B434}, 347 (1998).

\bibitem{PG} P. Galison, Nucl. Phys. {\bf B232}, 26 (1984).

\bibitem{GM} H. Georgi and M. Machacek, Nucl. Phys. {\bf B262}, 463 (1985);
R. Chivukula and H. Georgi, Phys. Lett. {\bf B182}, 181 (1986).

\bibitem{CG} M. Chanowitz and M. Golden, Phys. Lett. {\bf B165}, 105 (1985).

\bibitem{jack} 
J. Gunion, R. Vega, and J. Wudka, Phys. Rev. {\bf D42}, 1673 (1990);
Phys. Rev. {\bf D43}, 191 (1991).

\bibitem{andrew}
A. Akeroyd, Phys. Lett. {\bf B442}, 335 (1998); Phys.Lett. {\bf B353}, 
519 (1995).

\bibitem{future}
K. Cheung and D.K. Ghosh, work in progress.

\bibitem{logan}
H. Haber and H. Logan,Phys. Rev. {\bf D62}, 015011 (2000).

\bibitem{kundu2}
A. Kundu and B. Mukhopadhyaya, Int. J. Mod. Phys. {\bf A11}, 5221 (1996).

\bibitem{ciu}
M. Ciuchini, G. Degrassi, P. Gambino, and G. Giudice, Nucl. Phys. {\bf B527},
527 (1998).

\bibitem{kundu1}
D. Chakraverty and A. Kundu,  Mod. Phys. Lett. {\bf A11}, 675 (1996).


\bibitem{cteq}
CTEQ Collaboration (H. Lai et al.), Eur. Phys. J. {\bf C12}, 375 (2000).

\bibitem{han}
V. Barger and T. Han, Phys. Lett. {\bf B212}, 117 (1988).

\bibitem{multi-W}
F. Berends, H. Kuijf, B. Tausk, and W. Giele, Nucl. Phys. {\bf B357}, 
32 (1991).

\bibitem{tev2000}
{\it Future Electroweak Physics at the Fermilab Tevatron: Report of the
tev-2000 Study Group}, FERMILAB-PUB-96/082, Edited by D. Amidei and 
R. Brock.

\bibitem{spira}
M.~Spira, Fortsch.~Phys.~{\bf 46}, 203 (1998); 
T. Han and S. Willenbrock, Phys.~Lett.~{\bf B273}, 167 (1991).

\bibitem{madevent}
F. Maltoni and T. Stelzer, hep-ph/0208156.

\end{thebibliography}
\end{document}